\documentclass[11pt,twoside]{atmp}
\usepackage{amsmath,amssymb}

\usepackage[all]{xy}
\usepackage{color}

\usepackage{bm}
\def\R{\mathbb{R}}
\def\C{\mathbb{C}}

\def\sgn{{\rm sgn}}

\theoremstyle{definition}

\begin{document}

\title
{Boundary states and edge currents for free fermions}

\arxurl{math-ph/0808.1497}

\author[M.\ Leitner, W.\ Nahm]{Marianne Leitner, Werner Nahm}

\address{Dublin Institute for Advanced Studies\\
School of Theoretical Physics \\
10 Burlington Road, Dublin 4, Ireland}  
\addressemail{leitner@stp.dias.ie, wnahm@stp.dias.ie}



\date{\today}

\begin{abstract}
We calculate the ground state current densities for 2+1 dimensional free fermion theories with local, 
translationally invariant boundary states. Deformations of the bulk wave functions close to the edge and boundary states
both may cause edge current divergencies, which have to cancel in realistic systems. This yields restrictions on 
the parameters of quantum field theories which can arise as low energy limits of solid state systems. Some degree of
Lorentz invariance for boosts parallel to the boundary can be recovered, when the cutoff is removed.
\end{abstract}

\maketitle

\section{Introduction}

The prediction of a relativistically invariant quantum field theory in 2+1 dimensions as low energy
limit of a solid state system \cite{Semenoff:1984} was confirmed by investigation of graphene \cite{Novoselov:2005}. Modification of the
graphene structure may allow to obtain more general quantum field theories \cite{Haldane:1988}. Since solid state systems
typically have edges, the study of quantum field theories with boundaries acquired additional interest.
An important tool in this context is the bulk-boundary OPE. In particular, some of its singularities 
yield divergent observables and constitute obstructions to any experimental realization. The coefficient function 
of the identity in such an OPE is just the corresponding vacuum expectation value in the presence of the boundary.
Integrated expectation values of current operators can be interpreted as measurable charge or spin currents and
should remain finite in realistic systems. Here we study this obstruction in the simplest case of free fermions,
where vacuum expectation values can be interpreted as integrated contributions of the particles in the Dirac sea.

It is of course well known that such effects can prevent the lifting of a quantum mechanical system
to a quantum field theory, even without the presence of a boundary. In our case the Dirac sea of a single fermion yields 
a half-integral value of the Hall conductivity \cite{Jackiw:1984}, \cite{Redlich:1984}, \cite{Leitner:2007}. 
Consequently one needs an even number of fermions, though some of them 
may be shifted to infinite mass, which results in a Chern-Simons term instead. For boundary states 
we shall find more intricate obstructions, which depend on the numerical parameters of the boundary state.

\section{Boundary conditions from self-adjoint extensions}
\noindent
We consider spinor fields $\psi$ acted upon by the Dirac operator
\begin{equation}\label{Dirac-op}
H:=-i\sigma_1\frac{\partial}{\partial x}-i\sigma_2\frac{\partial}{\partial y}+m\sigma_3
\end{equation}
on the open half plane $\{\mathbf{x}=(x,y)\in\R^2|\:x>0\}$, as in \cite{GruLei:2006}.
This operator is a closable symmetric operator on the domain ${\mathcal{D}}(H)=C_0^\infty(\R_+\times\R,\C^2)$ 
of smooth functions with compact support vanishing in a neighbourhood of $x=0$. The possible boundary conditions are classified by its 
self-adjoint extensions. The latter can be obtained by the standard von Neumann theory. Here we give a modification, which is less 
general but better adapted to the present problem.
Since ${\mathcal{D}}(H)$ is dense in the Hilbert space ${\mathcal{H}}:=L^2(\R_+\times\R)\otimes\C^2$, the image of an arbitrary vector 
$\psi\in\mathcal{H}$ in the domain of a self-adjoint extension is uniquely determined by 
the unextended operator, so the problem is reduced to the determination of such domains. Let $H^*$ be the adjoint of the unextended $H$. 
By von Neumann theory, every self-adjoint extension of $H$ corresponds 1-1 to an isometry 
$V:\,N_+\rightarrow N_-$ between the defect spaces 
\begin{align*}
N_{\pm}=\{\psi\in{\mathcal{D}}(H^*)|\,H^*\psi=\pm i\mu\psi\}, 
\end{align*}
where $\mu$ is a fixed but arbitrary positive number. Indeed, $V$ gives rise to the self-adjoint extension $H_V$ with domain
\begin{align}\label{extended domain}
{\mathcal{D}}(H_V)={\mathcal{D}}(\bar{H})+(I-V)\,N_+. 
\end{align}
The choice $\mu=1$ is standard, but we shall see that the limit of large $\mu$ allows a more direct physical interpretation.
Applied to (\ref{Dirac-op}), $N_{\pm}$ is given by $k$ integrals over elements of the one-dimensional spaces 
\begin{align*}
N_{\pm}(k)
=
\C\:e^{-\lambda x+iky}
\begin{pmatrix}
1\\s_k^{\pm}
\end{pmatrix}
(1+O(\mu^{-1})),
\end{align*}
where
\begin{align*}
s_k^{\pm}
=\frac{i(k+\lambda)}{m\pm i\mu},
\quad\text{and}\quad
\lambda=\sqrt{\mu^2+k^2+m^2}. 
\end{align*}
For $\mu\rightarrow\infty$ the defect spaces consist of wave functions which become supported on an arbitrarily small neighborhood of the
boundary and factorize in a universal term $\exp(-\mu x)$ and a spinor of the form
\begin{align*}
g(y)
\begin{pmatrix}
1\\\pm 1
\end{pmatrix}
\quad
\text{with}\quad g\in L^2(\R).
\end{align*}
Thus isometries $N_+\rightarrow N_-$ become tensor products of a unitary map $V:L^2(\R)\rightarrow L^2(\R)$ with the map between
one dimensional subspaces of $\C^2$ which takes $(1,-1)$ to $(1,1)$. 
From (\ref{extended domain}) we read off that any
 $\psi\in{\mathcal{D}}(H_V)$ satisfies
\begin{align*}
\psi_1|_{x=0}&=(1-V)g,\\
\psi_2|_{x=0}&=(1+V)g. 
\end{align*}
If $1$ is not an eigenvalue of $V$, then  
\begin{align}\label{b.c.G}
\psi_2|_{x=0}=i\Gamma\:\psi_1|_{x=0}, 
\end{align}
where $\Gamma$ is the self-adjoint operator
\begin{align*}
\Gamma=-i(1+V)(1-V)^{-1}.
\end{align*}
The physical necessity of such a boundary condition can be understood as follows. Consider the conserved current densities
\begin{equation*}
j^{\mu}(\mathbf{x})
=\lim_{\mathbf{x}'\rightarrow\mathbf{x}}\psi(\mathbf{x}')^{\dagger}\,\sigma^\mu\,\psi(\mathbf{x}),.
\end{equation*}
One needs a boundary condition which implies $\int j^1(0,y)\:dy=0$, and this indeed follows from (\ref{b.c.G}).

We only study local boundary conditions, for which $j^1(0,y)$ vanishes identically and also impose
translational invariance. Thus $\Gamma$ has to commute with $d/dy$ and with multiplication by
functions of $y$. This implies that it is a real constant $\gamma$, and we obtain the boundary condition
\begin{equation*}
\psi_2(0,y)=i\gamma\,\psi_1(0,y).
\end{equation*}
which we shall use in the sequel. To include the case $V=1$, we consider $\gamma$ as an element of the projective real line, 
including the value $\gamma=\infty$.

For any $\gamma$ the Hamiltonian commutes with the PT transformation $\psi(x,y)\mapsto \sigma_3\psi^*(x,-y)$. The space reflection
$(x,y)\mapsto (x,-y)$ lifts by $\psi(x,y)\mapsto \sigma_1\psi(x,-y)$ to a map between systems with parameters $(m,\gamma)$ and $(-m,-\gamma^{-1})$. Because of this duality we only need to consider positive $m$. Arbitrary values of $m$ will only be considered at the very end of the calculation. It turns out that the results allow a continuous extension to
$m=0$, which is the case relevant for graphene.

The CPT map $\psi(x,y) \mapsto \sigma_2\psi(x,-y)$ relates the positive energy states for $(m,\gamma)$ to
the negative energy states for $(m,\gamma^{-1})$. A different type of duality is given by $\psi(x,y)\mapsto \sigma_2\psi(-x,y)$,
$(m,\gamma)\mapsto (-m,\gamma^{-1})$, which relates boundary states of complementary half planes.

\section{Eigenfunctions}

The Dirac operator commutes with $-i\partial/\partial y$ and consequently with
$$(i\partial/\partial x)^2 = H^2+(\partial/\partial y)^2-m^2.$$
We denote the real eigenvalues of these hermitean operators by $k$ and $l^2$. We take $l\in\R^+$ for positive $l^2$ 
and $l=i\lambda$ with $\lambda\in\R^+$ for negative $l^2$. The corresponding eigenvalues $E$ of $H$ satisfy $E^2=k^2+l^2+m^2$.

According to the sign of $l^2$ the spectrum of $H$ has two parts, which will be refered to as bulk and edge. 
Bulk eigenfunctions $u_{lk}$ with given $k,l^2$ have the form
$$ u_{lk}(x,y) = \exp(ilx+iky)\chi_++\exp(-ilx+iky)\chi_-$$ 
with constant spinors $\chi_\pm$. The equation $Hu_{lk}=Eu_{lk}$ and the boundary condition reduce to 
$(m-E,\,\pm l-ik)\chi_\pm=0$ and
$(\chi_++\chi_-)_2=i\gamma (\chi_++\chi_-)_1$. For each $l>0$ there is a unique solution up to normalization, such that for the bulk part
we just recover the spectrum of the Dirac operator on the whole plane, with a gap $\Delta=(-m,m)$.
We normalize these eigenfunctions with respect to the measure $dk\,dl/(2\pi^2)$ and choose phases such that

\begin{equation*}
u_{lk}(x,y)
=\left(e^{i\phi(l,k)}
\begin{pmatrix}
i\rho_{lk}\\ 1
\end{pmatrix}
e^{ilx+iky}
-
\begin{pmatrix}
i\rho^*_{lk} \\ 1
\end{pmatrix}
e^{-ilx+iky}\right)\sqrt{\frac{E-m}{4E}},
\end{equation*}
where
\begin{equation}\label{phi}
e^{i\phi(l,k)}=\frac{1+\gamma\rho_{lk}^*}{1+\gamma\rho_{lk}}
\end{equation}
with 
\begin{equation}\label{rho}
\rho_{lk}:=\frac{k+il}{m-E}.
\end{equation}

Edge eigenfunctions \cite{GruLei:2006} have the form $U_k=\exp(-\lambda x+iky)\chi$. 
We normalize them with respect to the measure $dk/\pi$ and choose phases such that
\begin{equation*}
U_k(x,y)
=\sqrt{\frac{\lambda}{1+\gamma^2}}
\begin{pmatrix}
i\\ -\gamma
\end{pmatrix}
e^{-\lambda x+iky}. 
\end{equation*}
The equation $HU_k=EU_k$ yields
\begin{equation*}
E =\frac{2\gamma}{1+\gamma^2}k+\frac{1-\gamma^2}{1+\gamma^2}m
\end{equation*}
and
\begin{equation*}\label{lambda}
\lambda 
=\frac{\gamma^2-1}{\gamma^2+1}k+\frac{2\gamma}{\gamma^2+1}m.
\end{equation*}
Thus $E,\lambda$ are one-valued functions of $k$, and for any $k$ one has a unique edge state whenever $\lambda>0$
and no edge state otherwise. The latter inequality restricts the possible values of $k,E$. The limiting values for
$\lambda=0$ lie on the mass shell. When we put
\begin{equation*}
 \eta=\sgn\frac{1-\gamma}{1+\gamma},
\end{equation*}
then positive energy half of the mass shell is touch for $m\eta=1$ and the negative energy half for $m\eta=-1$.
One finds energies within the spectral gap $\Delta$ of the bulk iff $m\gamma>0$. 

The linear dispersion relation for the boundary equation implies that edge states travel at the fixed velocity

$$v_{\mathit edge}=\frac{2\gamma}{1+\gamma^2}.$$
Note that $v_{\mathit edge}$ determines $\gamma$ up to the replacement of $\gamma$ by $\gamma^{-1}$, which is the one induced by CPT.
The latter transformation changes the signature $\eta$. Altogether, the boundary condition can be characterized by 
$v_{\mathit edge}$ and $\eta$, except for the fact that $\eta$ is undefined in the limiting cases $\gamma=\pm 1$. 
One observes that
\begin{enumerate}
\item 
for vanishing edge velocity, i.e. for $\gamma=0$ and $\gamma=\infty$, edge states have $E=m$ and $E=-m$, respectively, independently of $k$.
This case is probably unphysical, since all edge states have the same energy and the thermodynamic partition function diverges for a system 
of infinite width. Since the width of the edge excitation increases linearly with $k$, the thermodynamic properties of a system of a finite 
system would depend on the sample size.
\item
The edge velocity is equal to the maximal bulk velocity for the CPT invariant boundary conditions $\gamma=\pm 1$. These yield $E=\gamma k$ 
and $\lambda=\gamma m$ independently of $k$. For $\gamma m< 0$  there are no edge states, for $\gamma m >0$ all edge states have the same 
dependence on $x$. This yields particular divergences, so we will tacitly exclude $\gamma=\pm 1$ in most of the subsequent calculations.
\end{enumerate}

In $(k,E)$-space the half-line of edge states is tangent to the half-hyperboloid given by $l\geq 0$ and $E^2=k^2+m^2+l^2$ at $l=\lambda=0$.
For $\gamma^2<1$ this happens at positive energy and for $\gamma^2>1$ at negative energy.
The tangency has consequences for the behaviour at large distance from the boundary, as we shall see later.

The portion of edge energies within $\Delta$ gives rise to an edge current parallel to the boundary, 
with edge-conductivity equal to (in units of $\frac{e^2}{h}$) \cite{GruLei:2006}

\begin{equation}\label{Gl: result for sigma}
\sigma^{\text{edge}}
=\begin{cases} \sgn(m), & \text{if }m\gamma>0, \\
                  \quad 0, & \text{otherwise}.
 \end{cases}
\end{equation}
This applies whenever some subinterval of the gap is actually occupied, independently
of the size of this subinterval. 

\section{The current densities}

When no boundary exists, Fermi energies within the gap $\Delta$ obviously yield the same theory.
This means that any pair $(\gamma, E_F)$ with $E_F\in\Delta$ yields a boundary state of the standard
Lorentz invariant free fermion theory. We want to calculate the corresponding expectation values of the
current densities. The bulk contribution does not depend on the choice of $E_F$ and edge
contributions for energies within $\Delta$ are finite and were calculated in \cite{GruLei:2006}. Thus it
suffices to consider $E_F=-m$.

The required vacuum expectation values are
\begin{equation*}
\left< j^{\mu}\right> = j^{\mu}_{\text{bulk}}+j^{\mu}_{\text{edge}},
\end{equation*}
where
\begin{align*}
j^{\mu}_{\text{bulk}}&= \int j^{\mu}_{lk}|_{\text{bulk}}\,\frac{dl\,dk}{2\pi^2},\\
j^{\mu}_{\text{edge}}&= \int \Theta(-m-E)\,j^{\mu}_k|_{\text{edge}}\,\frac{dk}{\pi},
\end{align*}
and
\begin{align*}
j^{\mu}_{lk}(\mathbf{x})|_{\text{bulk}}
&=u_{lk}^\dagger(\mathbf{x})\,\sigma^\mu\,u_{lk}(\mathbf{x}),\\
j^{\mu}_k(\mathbf{x})|_{\text{edge}}
&=U_k^\dagger(\mathbf{x})\,\sigma^\mu\,U_k(\mathbf{x}).
\end{align*}
Here the variable $E$ in $u_{lk}$ is the negative roots of $k^2+l^2+m^2$.

Note that $j^1_{lk}|_{\text{bulk}}$ and $j^1_k|_{\text{edge}}$ vanish identically. 
On the other hand eqs. (\ref{phi}, \ref{rho}) yield
\begin{equation}\label{j2-bulk}
j^2_{lk}(\mathbf{x})|_{\text{bulk}}=\frac{k}{E}-\frac{1}{E}\Re\left(\frac{f}{g}\,e^{-2ilx}\right).
\end{equation}
where
\begin{align*}
g&=m-E+\gamma(k-il),\\
 f&=(k-il)g^*.
\end{align*}
The first term on the r.h.s. of eq. (\ref{j2-bulk}) is an odd function of $k$, so that its contribution to the vacuum 
expectation value of $j^2$ vanishes for any symmetric regularization. For the second term we also will find a partial cancelation.

We substitute $k\in(-\infty,\infty)$ by $v=e^{arcsinh(k/a)}\in (0,\infty)$, which
yields for negative $E$
\begin{equation}\label{dv}
-\frac{dk}{E}=\frac{dv}{v}.
\end{equation}
Moreover
\begin{equation*}
-\frac{f}{g}\frac{k}{E}
=\left(\frac{a^2}{4}(v-v^{-1})^2+l^2\right)(D_1^{-1}+ D_2^{-1})\:dv.
\end{equation*}
Here
\begin{align*}
D_1&=\frac{a}{2}(\gamma+1)(v-v_3)(v-v_4),\\
D_2&=\frac{a}{2}(\gamma^{-1}+1)(v-v_3)(v+v_4),
\end{align*}
where
\begin{equation*}
a=\sqrt{l^2+m^2}
\end{equation*}
and
\begin{align*}
v_3&=\frac{il+m}{\sqrt{l^2+m^2}}\frac{\gamma-1}{\gamma+1},\\
v_4&=\frac{-il+m}{\sqrt{l^2+m^2}}.
\end{align*}
The two denominators $D_1$ and $D_2$ are formally related by a reflection across the boundary, with
$(m,\gamma)\mapsto (-m,\gamma^{-1})$.

One obtains a partial fraction expansion
\begin{equation*}
-\frac{f}{g}\frac{dk}{E}
=\left\{P_1(v)+P_2(v)+P_3(v)+P_4(v)\right\}\:dv
\end{equation*}
with
\begin{align*}
P_1(v)&=\frac{a}{2},\\
P_2(v)&=-\frac{a}{2}v^{-2}.
\end{align*}

To regularize the momentum integrals we have to introduce a cutoff. In a solid state context the underlying physics
selects a stationary Lorentz frame, in which a symmetric cutoff in $k$ is natural. For this
cutoff the $dv$ integrals of these two terms cancel against each other due to the
symmetry $k\mapsto -k$, $v\mapsto 1/v$. The remaining terms are
\begin{align*}
P_3(v)=&il\frac{\gamma+1}{\gamma-1}v^{-1},\\
P_4(v)=-&il\frac{4\gamma}{\gamma^2-1}(v-v_3)^{-1}.
\end{align*}
For a symmetric cutoff $\Lambda$ of the $v$-integration one finds
\begin{equation*}
\int \Re(P_3(v)e^{-2ilx}) \frac{dl dv}{2\pi^2}=
-\frac{1}{2\pi}\frac{\gamma+1}{\gamma-1}\delta'(x)\ln\Lambda.
\end{equation*}

To calculate the $P_4$ integral we need
\begin{equation*}
\int\left(v-\frac{il+m}{\sqrt{l^2+m^2}}\frac{\gamma-1}{\gamma+1}\right)^{-1}\:dv
=\ln(\Lambda)-\ln\left(\frac{1+\gamma}{1-\gamma}\sqrt{\frac{m+il}{m-il}}\right).
\end{equation*}
where the principal value of the logarithm has to be taken, so that
\begin{equation*}
 \ln\left(\frac{1+\gamma}{1-\gamma}\sqrt{\frac{m+il}{m-il}}\right)
=\ln\left(\sqrt{\frac{m+il}{m-il}}\right)+\ln\left|\frac{1+\gamma}{1-\gamma}\right|-\pi i\,\Theta(\gamma^2-1).
\end{equation*}

To calculate the $dl$ integral over the logarithmic part we use
\begin{equation*}
2\int_0^{\infty}il\ln\left(\sqrt{\frac{m+il}{m-il}}\right)cos(2lx)\:dl
=\int_{-\infty}^{\infty}il\ln\left(\sqrt{\frac{m+il}{m-il}}\right)e^{2ilx}\:dl.
\end{equation*}
The argument of $\ln$ has branching points at $l=\pm im$. We may move the integration path towards the cut with branching point $l=im$.
Since the discontinuity of the logarithm is $i\pi$, this yields an elementary integral. Altogether we obtain 
\begin{align}\label{j2bulk}
j^2&(\mathbf{x})_{\text{bulk}}\\
=&-\frac{1}{2\pi}\frac{\gamma^2+1}{\gamma^2-1}\delta'(x)\,\ln\Lambda \quad +\frac{\gamma}{\pi(\gamma^2-1)}\ln\left|\frac{1+\gamma}{1-\gamma}\right|\,\delta'(x)\nonumber\\
&+\frac{\gamma}{2\pi(\gamma^2-1)}\left(\frac{1}{2x^2}+\frac{m}{x}\right)e^{-2m x}
-\frac{\gamma}{\pi(\gamma^2-1)}\frac{1}{2x^2}\,\Theta(\gamma^2-1)\nonumber.
\end{align}

For the contribution from the edge functions, we have according to (\ref{lambda}) 
\begin{equation*}
dk\:\frac{\gamma^2-1}{\gamma^2+1}=d\lambda.
\end{equation*}
We have to integrate over those $\lambda$ for which $\lambda>0$ and $E<-m$. This yields
\begin{equation*}
j^2(\mathbf{x})_{\text{edge}}\\
=\frac{\gamma}{\pi(\gamma^2-1)}\left[\frac{1}{2x^2}\Theta(\gamma^2-1)-
\left(\frac{1}{2x^2}+\frac{m}{\gamma x}\right)e^{-2m x/\gamma}
\Theta(\gamma)\right].
\end{equation*}
Note that $j^2_{\text{edge}}$ vanishes for $\gamma\in(-1,0)$ and is non-singular for $\gamma>1$.

For $\gamma^2>1$ both the bulk and the edge current densities decay algebraically at large distance from the boundary.
In their sum these terms cancel, an interesting effect which reflects the tangency of the edge state and bulk state
manifolds in $(k,E)$ space, which was discussed in the previous section. 

In all cases the singular part of the vacuum expectation value has the form
\begin{equation*}
 \left< j^2(\mathbf{x})\right>_{\text{singular}}=-\frac{1}{2\pi}\frac{\gamma^2+1}{\gamma^2-1}\delta'(x)\,\ln\Lambda 
-\frac{|\gamma|}{4\pi(\gamma^2-1)}\frac{1}{x^2}.
\end{equation*}
Remarkably, this equation is invariant under $y$ reflection together with $\gamma\mapsto -\gamma^{-1}$,
such that it remains true for $m<0$ and is in fact independent of $m$. The regular part of the expectation
value is continuous in $m$ and vanishes for $m=0$. For infinite $m$ the contribution becomes localized at the
boundary and is given by the $\delta'$ terms in (\ref{j2bulk}). It may be interesting to rederive this result
from a Chern-Simons Lagrangian.

\section{Conclusion}
A single free fermion yields a divergent edge current on a half plane. For $N$ fermions with boundary
conditions $\gamma_n$ the cut-off dependent edge current vanishes for
\begin{equation*}
 \sum_{n=1}^N\frac{\gamma_n^2+1}{\gamma_n^2-1}=0.
\end{equation*}
One still has a dipolar edge current
\begin{equation*}
\sum_{n=1}^N \frac{\gamma_n}{\pi(\gamma_n^2-1)}\ln\left|\frac{1+\gamma_n}{1-\gamma_n}\right|\,\delta'(x),
\end{equation*}
which in general does not vanish. In addition there is a $1/x^2$ singularity, which yields an unphysical
divergent particle transport in the neighborhood of the edge, unless
\begin{equation*}
\sum_{n=1}^N\frac{|\gamma_n|}{\gamma_n^2-1}=0. 
\end{equation*}

Since the bulk system and the boundary at $x=0$ are Lorentz invariant under boosts in the $y$ direction, one should be able to act with
Lorentz transformations on the boundary conditions. In the context of relativistc quantum mechanics this is straightforward.
The action on the parameters $\gamma_n$ can be determined from eq. (3) or more easily from eqs. (8, 9), 
according to which boundary states are characterized by a Lorentz invariant signature $\eta_n$ and their fixed travel velocity

$$v_n=\frac{2\gamma_n}{1+\gamma_n^2}.$$
This translates into a rapidity $\theta_n$ given by $\tanh\theta_n=v_n$, or equivalently by
$$\eta_n\exp\theta_n = \frac{1+\gamma_n}{1-\gamma_n},$$
to which the rapidity of the Lorentz boost is added. When one moves from one-particle states
to field theory, control of divergencies requires a cutoff which breaks Lorentz invariance. It may or may not be restored, when this cutoff 
is taken to infinity. One would expect that a restoration of Loretz invariance implies CPT invariance. Due to the unbroken PT invariance
this should yield invariance under charge conjugation and consequently an absence of net boundary currents.
In our case we will find an unexpectedly subtlety, however. In terms of edge velocities and signatures the two divergency 
cancellation conditions can be rewritten as
$$\sum_{n=1}^N \eta_n\exp(\epsilon_n\theta_n) =0= \sum_{n=1}^N \eta_n\exp(-\epsilon_n\theta_n),$$
where 
$$\epsilon_n=\sgn v_n.$$
For $N=1$ there is no solution, in agreement with the restriction to an even $N$ coming from the bulk physics.
For $N=2$ one needs $\gamma_2=-\gamma_1^{-1}$, thus invariance under charge conjugation. More generally, pairs of edge states related by 
charge conjegation do not contribute to the breaking of Lorentz invariance or to net boundary currents, as expected. There is an
unexpected remainder of Lorentz invariance in more general situation, however. When all $v_n$ of the remaining states have the same sign, 
the cancellation conditions are invariant under sufficiently small boosts which do
not change the sign of edge velocities. When $v_n$ of different signs occurs, no aspect of Lorentz invariance is recovered when the cutoff
is removed. In any case, breaking of Lorentz invariance by edge states does not seem to be an argument against the realization of such 
systems in solid state physics, where this invariance is only a low energy phenomenon.

It will be interesting to see to what extent the $\gamma$ values can be controled experimentally.
Systems without time reversal invariance would be particularly interesting, but are certainly
very difficult to realize.

\end{document}